\documentclass[a4paper,12pt]{article}
\usepackage[pctex32]{graphics}

\newcommand{\sect}[1]{\setcounter{equation}{0}\section{#1}}

\begin{document}
\topmargin 0pt \oddsidemargin 0mm
 
\renewcommand{\thefootnote}{\fnsymbol{footnote}}
\begin{titlepage}
\begin{flushright}
INJE-TP-03-06\\
hep-th/0306180
\end{flushright}
 
\vspace{5mm}
\begin{center}
{\Large \bf Role of the cosmological constant in the holographic
description of the early universe} \vspace{12mm}
 
{\large  Yun Soo Myung\footnote{e-mail
 address: ysmyung@physics.inje.ac.kr}}
 \\
\vspace{10mm} {\em  Relativity Research Center and School of
Computer Aided Science, Inje  University Gimhae 621-749, Korea}
\end{center}
 
\vspace{5mm} \centerline{{\bf{Abstract}}}
 \vspace{5mm}
We investigate the role of the cosmological constant  in the
holographic description of a radiation-dominated universe
$C_2/R^4$ with a positive cosmological constant $\Lambda$. In
order to understand the nature of cosmological term, we first
study the newtonian cosmology. Here we find two aspects of the
cosmological term:  entropy ($\Lambda \to S_{\rm \Lambda}$) and
energy ($\Lambda \to E_{\rm \Lambda}$). Also we solve the
Friedmann equation parametrically to obtain another role. In the
presence of the cosmological constant, the solutions are described
by the Weierstrass elliptic functions on torus and have modular
properties. In this case one may expect to have a two-dimensional
Cardy entropy formula but the cosmological constant plays a role
of  the modular parameter $\tau(C_2,\Lambda)$ of torus.
Consequently the entropy concept of the cosmological constant is
very suitable for establishing  the holographic entropy bounds in
the early universe. This contrasts to the  role of the
cosmological constant as a dark energy in the present universe.

\end{titlepage}
 
\newpage
\renewcommand{\thefootnote}{\arabic{footnote}}
\setcounter{footnote}{0} \setcounter{page}{2}
 
%%===================section 1 ====================
\sect{Introduction}
 
Nowadays the cosmological constant plays an important role in
several fields: cosmology, astronomy, particle physics and string
theory. The reason is twofold. One is that the inflation turned
out to be a successful tool to resolve the problems of the hot big
bang model~\cite{Infl}. Thanks to the recent observations of the
cosmic microwave background anisotropies and large scale structure
galaxy surveys, it has become widely accepted by the cosmology
community~\cite{JGB}. The idea of primordial inflation is based on
the very early universe dominance of vacuum energy density of a
hypothetical scalar field, the inflaton. This produces the
quasi-de Sitter spacetime~\cite{Hogan} and during the slow-roll
period, the equation of state can be approximated by the vacuum
state as $p\approx -\rho$ like $p_{\rm \Lambda}=\omega\rho_{\rm
\Lambda}, \omega=-1$ for the cosmological constant
$\Lambda$~\cite{FKo}. The other is that an accelerating universe
(with positive cosmological constant) has recently proposed to
interpret the astronomical data of supernova. In this case, the
cosmological constant has been identified with a dark exotic form
of energy that  is smoothly distributed and which contributes 2/3
to the critical density of the present universe\footnote{Recently,
the dark form of energy is classified according to the equation of
state : quintessence with $-1 <\omega<-1/3$, cosmological constant
with $\omega=-1$, and phantom energy with $\omega
<-1$~\cite{phan}.}.

On the other hand  we have to build cosmology from the quantum
gravity for completeness, but now we are far from it. Although we
are lacking for a complete understanding of the quantum gravity,
there exists the holographic principle.
 This principle  is mainly based on  the idea that for a given volume
$V$,
  the state
 of maximal entropy is given by the largest black hole that fits inside
$V$.
 't Hooft and Susskind~\cite{Hooft} argued that the microscopic entropy
$S$
 associated with the volume $V$ should be less than the
Bekenstein-Hawking
entropy:  $S \le A/4G$ in the units of $c=\hbar=1$~\cite{Beke}.
Here the horizon area $A$ of a black hole equals the surface area
of the boundary of $V$. That is,  if one reconciles quantum
mechanics and gravity, the observable degrees of freedom of our
three-dimensional universe comes from a two-dimensional surface.
Actually holographic area bounds limit  the number of physical
degrees of freedom in the bulk spacetime.

The implications of the holographic principle  for the early
universe have been investigated in the  literature. Following an
earlier work by Fischler and Susskind~\cite{FS} and works
in~\cite{Hubb,Bous}, it was argued  that the maximal entropy
inside the universe is given by the {\bf Hubble entropy}. This
geometric entropy plays an important role in establishing the
cosmological holographic principle in the early universe. Roughly
speaking, the total matter entropy should be less than or equal
the Bekenstein-Hawking entropy of the Hubble-size black hole
($\approx H V_{\rm H}/4G_{n+1}$) times the number ($N_{\rm
H}\approx V/V_{\rm H}$) of Hubble regions in the early universe.
That is, the Hubble entropy as an upper bound on the total matter
entropy is proportional to $HV/4G_{n+1}$. Furthermore, Verlinde
fixed the prefactor as $(n-1)$ and
 proposed the new holographic bounds Eq.(\ref{4eq5}) in
a radiation-dominated phase by introducing three
entropies~\cite{Verl}: Bekenstein-Verlinde entropy ($S_{\rm BV}$),
Bekenstein-Hawking entropy ($S_{\rm BH}$), and Hubble entropy
($S_{\rm H}$). As an  example, such a radiation-dominated phase is
provided by a conformal field theory (CFT) with a large central
charge which is dual to the AdS-black hole~\cite{SV}. In this case
it appeared  an interesting relationship between the Friedmann
equation governing the cosmological evolution and the square root
form of entropy-energy relation, called Cardy-Verlinde
formula~\cite{Cardy}. Although the Friedmann equation has a
geometric origin and  the Cardy-Verlinde formula is designed only
for the matter content, it  suggested  that both  may arise from a
single  fundamental theory.  However, this approach remains
obscure for a radiation-dominated universe with a positive
cosmological constant~\cite{CM1}. This is mainly due to the
unclear role of the cosmological constant in the  holographic
description of the early universe.
 
In this work we will clarify the role of the cosmological term in
the early universe. For this purpose we introduce the newtonian
cosmology and the parametric solution to the Friedmann equation.
We will show that the geometric entropy interpretation of the
cosmological term plays an important role  in establishing the
holographic entropy bound for a radiation-dominated universe with
a positive cosmological constant. Finally we wish to point out the
different roles  of the cosmological constant in the early
universe and in the present universe.
 
 The relevant equation is an $(n+1)$-dimensional
Friedmann-Robertson-Walker (FRW) metric with $k=1$
\begin{equation}
\label{1eq1} ds^2 =-dt^2 +R(t)^2  d\Omega^2_{n} ,
\end{equation}
where $R$ is the  scale factor of the universe and $d\Omega^2_{n}$
denotes the line element of an $n$-dimensional unit sphere.  A
cosmological evolution is determined by the two Friedmann
equations
\begin{eqnarray}
\label{1eq2}
 && H^2 =\frac{16\pi G_{n+1}}{n(n-1)}\frac{E}{V}
-\frac{1}{R^2}
     +\frac{1}{l^2_{n+1}},  \\
     \label{1eq3}
&& \dot H =-\frac{8\pi G_{n+1}}{n-1}\left (\frac{E}{V} +p\right)
    +\frac{1}{R^2},
\end{eqnarray}
where $H$ represents the Hubble parameter with the definition
$H=\dot R/R$ and the overdot stands for  derivative with respect
to the cosmic time $t$,  $E$ is the total energy of matter filling
the universe, and $p$ is its pressure. $V$ is the volume of the
universe, $V=R^n \Omega_n$ with $\Omega_n$ being the volume of an
$n$-dimensional unit sphere, and $G_{n+1}$ is the newtonian
constant in ($n+1$) dimensions. Here we assume the equation of
state for any matter:
 $p=\omega \rho,~ \rho=E/V$. For our purpose, we include
 the curvature radius of de Sitter space $l_{n+1}$
 which relates to the cosmological constant
 via $1/l^2_{n+1}=2\Lambda_{n+1}/n(n-1)$. For $n=3$ case, we use
 the notation of $G,\Lambda$ instead of $G_4,\Lambda_4$.
 
The organization of this paper is as follows. In section 2, we
study the newtonian cosmology. Section 3 is devoted to solving the
Friedmann equation in a parametrical way to find out the role of
the cosmological term. The cosmological holographic  bounds for a
radiation-dominated universe without/with a positive cosmological
constant are discussed in section 4. Finally we discuss our
results in section 5.

%%====================section 2=====================
\sect{Newtonian cosmology} In order to understand the cosmological
term $\Lambda$ in the Friedmann equation, let us study the
newtonian cosmology in (3+1)dimensions. Even though the newtonian
cosmology is valid for the matter-dominated universe (that is, it
is non-relativistic), this approach is useful for understanding
the origin of the cosmological term. We propose that the universe
consists of a number of galaxies with their mass $m_i$ and
position ${\bf r}_i(t)=r_i(t)\hat {\bf r} $ as measured from a
fixed origin $O$. Then  the kinetic energy of the system $T$ is
given by
\begin{equation}
\label{2eq1} T= \frac{1}{2} \sum^{n}_{i=1} m_i \dot r^2_i.
\end{equation}
The total gravitational potential energy $V$ is
\begin{equation}
\label{2eq2} V_g= -G \sum^{n}_{i<j} \frac{m_i m_j}{|{\bf r}_i-{\bf
r}_j|}.
\end{equation}
Assuming  that there exists a cosmological force acting on the
$i$-th galaxy of the form ${\bf F}_i=\frac{\Lambda}{3} m_i{\bf
r}_i$ with a  constant $\Lambda$ leads to the cosmological
potential energy
\begin{equation}
\label{2eq3} V_c= -\frac{\Lambda}{6} \sum^{n}_{i=1} m_i r^2_i.
\end{equation}
Then the total energy $E$ of this system is given by
\begin{equation}
\label{2eq4} E=\frac{1}{2} \sum^{n}_{i=1} m_i \dot r^2_i -G
\sum^{n}_{i<j} \frac{m_i m_j}{|{\bf r}_i-{\bf
r}_j|}-\frac{\Lambda}{6} \sum^{n}_{i=1} m_i r^2_i.
\end{equation}
Suppose that the distribution and motion of the system is known at
some fixed epoch $t=t_0$. By the cosmological principle of
homogeneity and isotropy, the radial motion at any time $t$ is
then given by $r_i(t)=S(t)r_i(t_0)$ where $S(t)$ is a universal
function of time which is the same for all galaxies and is  called
the scale factor. Substituting this into Eq.(\ref{2eq4}) leads to
\begin{equation}
\label{2eq5} E=A \dot S(t)^2 -\frac{B}{S(t)} -DS(t)^2,
\end{equation}
where the coefficients are positive constants given by
\begin{equation}
\label{2eq6} A= \frac{1}{2} \sum^{n}_{i=1} m_i [r_i(t_0)]^2,~~ B=G
\sum^{n}_{i<j} \frac{m_i m_j}{|{\bf r}_i(t_0)-{\bf
r}_j(t_0)|},~~D=\frac{\Lambda}{6} \sum^{n}_{i=1} m_i
[r_i(t_0)]^2=\frac{\Lambda}{3} A.
\end{equation}
This is one form of the cosmological differential equation for the
scale factor $S(t)$. If the universe with $\Lambda=0$ is
expanding,  $A$-term decreases since the total energy remains
constant as $B$-term decreases. Therefore the expansion must slow
down. If $\Lambda$ is positive, all galaxies experience a cosmic
repulsion, pushing them away form the origin out to infinity. If
$\Lambda$ is negative, all galaxies experience a cosmic attraction
towards  the origin. Introducing a new scale factor with $R(t)=\mu
S(t)$, Eq.(\ref{2eq5}) takes the form\footnote{Similarly, assuming
the five-dimensional newton potential $V_{5g}=-G_5 \sum^{n}_{i<j}
\frac{m_i m_j}{|{\bf r}_i-{\bf r}_j|^2}$, one can find the
equation for a radiation-dominated universe in four-dimensional
spacetime as $\dot R^2=\frac{C_2}{R^2} + \frac{\Lambda}{3}R^2 -k$.
Even though it is a non-relativistic approach to obtain a
relativistic matter of radiation, this may provide us a hint for
 interpreting the cosmological term in the Friedmann equation.}
\begin{equation}
\label{2eq7} \dot R^2=\frac{C_1}{R} + \frac{\Lambda}{3}R^2 -k,
\end{equation}
where the constants $C_1$ and $k$ are defined by $C_1=B \mu^3/A$
and $k=-\mu^2 E/A$. When $E=0$, $\mu$ is arbitrary. However, if $E
\not=0$, one may choose $\mu^2=A/|E|$ so that  $k=1,0,-1$. This
equation is exactly the same form of the Friedmann equation of
relativistic cosmology. Although there exist ambiguities in
determining the cosmological parameters $C_1$ and $k$, one finds
that the cosmological term has a slightly different origin from
others. The  term in the left-hand side of Eq.(\ref{2eq7})
originates from the kinetic energy,  the first term (last term) in
the right-hand side come from the potential energy (total energy)
whereas  the second term from the constant cosmological repulsion
or attraction. We are interested in the role of the cosmological
term in the holographic description of cosmology. As are shown in
Eqs.(\ref{2eq5}) and (\ref{2eq6}), a shape of the cosmological
term is similar to the kinetic term which can be expressed as the
Hubble entropy. On the other hand its nature belongs to the
$B$-potential term that can be transformed into the energy term.
These two pictures will be used for confirming the cosmological
holographic bounds for a radiation-dominated universe with a
cosmological constant.

 \sect{Parametric cosmological solutions}
There  exists another approach to establishing the cosmological
holographic principle.  In this case, it seems that the
cosmological constant  plays a role of a parameter in deriving a
Cardy formula on torus. In this  section, we study this approach
to investigate a role of the cosmological term explicitly.
 
\subsection{Case  without a cosmological constant}
 
In general we have three cosmological parameters $C,\Lambda,k$.
Let us first consider the matter-dominated Friedmann equation with
$\Lambda=0,k=1$
\begin{equation}
\label{3eq1} \dot R^2=\frac{C_1}{R}  -1
\end{equation}
with $C_1= 8 \pi G \rho_{m0}/3$. Here the energy density for a
matter-dominated universe is given by
$\rho_m=E_m/V=\rho_{m0}/R^3$. Introducing an arc parameter $\eta$
(radians of arc distance on $S^3$), one finds the
solution~\cite{MTW}
\begin{equation}
\label{3eq2} R(\eta)=\frac{C_1}{2}(1-\cos
\eta),~~t(\eta)=\frac{C_1}{2}(\eta-\sin \eta).
\end{equation}
The range of $\eta$ from start of expansion to end of
recontraction is $2 \pi$ and the curve of $R(t)$ is cycloid. The
limiting form of law of expansion at the early times is given by
\begin{equation}
\label{3eq3} R \approx \frac{C_1}{4} \eta^2,~~t \approx
\frac{C_1}{12} \eta^3~~\to R \approx (9C_1/4)^{1/3}
t^{\frac{2}{3}}
\end{equation}
which is consistent with the solution to the matter-dominated
universe.
 
Now we consider the radiation-dominated Friedmann equation
\begin{equation}
\label{3eq4} \dot R^2=\frac{C_2}{R^2}  -1
\end{equation}
with $C_2= 8 \pi G \rho_{r0}/3$. The energy density for a
radiation-dominated universe is given by
$\rho_r=E_r/V=\rho_{r0}/R^4$. Introducing the same arc parameter
$\eta$, one finds the solution\footnote{This is the same form of
the entropy solution $ S_H( \eta)=S_{BV}\sin
\eta,~~S_{BH}(\eta)=S_{BV}(1-\cos \eta)$ to the circular relation
of the holographic entropies with $\Lambda=0$:
$S_H^2+(S_{BV}-S_{BH})^2=S_{BV}^2$~\cite{Verl}. Here $\eta$
corresponds to the conformal time coordinate via $R d
\eta=(n-1)dt$. $S_{\rm BV}$ is constant, $S_{\rm H}$ and $S_{\rm
BH}$ change with time.}
\begin{equation}
\label{3eq5} R(\eta)=\sqrt{C_2}\sin
\eta,~~t(\eta)=\sqrt{C_2}(1-\cos \eta).
\end{equation}
The range of $\eta$ from start of expansion to end of
recontraction is $\pi$ and the curve of $R(t)$ is semicircle.
 The limiting form of law of
expansion at the early times are given by
\begin{equation}
\label{3eq6} R \approx \sqrt{C_2} \eta,~~t \approx
\frac{\sqrt{C_2}}{2} \eta^2~~\to R \approx
2^{1/2}C_2^{1/4}t^{\frac{1}{2}}
\end{equation}
which leads to the well-known solution for the radiation-dominated
universe. The parametric solutions to the Friedmann equation with
$\Lambda=0$ are determined by the elementary trigonometric
functions. But their nature is different: one is cycloid and the
other is semicircle.

\subsection{Case with a cosmological constant}
We start with  the matter-dominated Friedmann equation with
$\Lambda \not=0,k=1$
\begin{equation}
\label{3eq7} \dot R^2=\frac{C_1}{R} +\frac{\Lambda}{3}R^2 -1.
\end{equation}
 Introducing an idea of elliptic curves on torus $T^2$,
 one finds the solution expressed in terms of the Weierstrass
 function as~\cite{KW}
\begin{equation}
\label{3eq8}
R(u,\tau)=\frac{3C_1}{12\wp(u+\epsilon,\tau)+1},~~t(u,\tau)=
\sqrt{\frac{3}{\Lambda}}
\Big[
\log\Big(\frac{\sigma(u+\epsilon-v_0)}{\sigma(u+\epsilon+v_0)}\Big)+2u
\zeta(v_0) \Big],
\end{equation}
where $\wp(z|\tau),\sigma(z|\tau),\zeta(z|\tau)$ are the
Weierstrass' family of functions: Weierstrass, Weierstrass sigma,
Weierstrass zeta functions, respectively. $u(C_1,\Lambda)$ is the
complex coordinate and $\tau(C_1,\Lambda)$ is a modular parameter.
These two  describing a torus are actually functions of both $C_1$
and $\Lambda$. $\epsilon$ is a constant of integration. The
Weierstrass function $\wp$ satisfies the equation of an elliptic
curve, a Riemann surface of genus 1 (torus)
\begin{equation}
\label{3eq9} (\wp')^2=4 \wp^3-g_2 \wp-g_3
\end{equation}
where the cubic invariants are given by
\begin{equation}
\label{3eq10} g_2=\frac{1}{12},~~g_3=\frac{1}{216}-\frac{\Lambda
C_1^2}{48}.
\end{equation}
Also it is a meromorphic modular form of weight 2 under ${\bf
SL}(2,Z)$ transformation,
\begin{equation}
\label{3eq11} \wp\Big(\frac{z}{c
\tau+d},\frac{a\tau+b}{c\tau+d}\Big)=(c\tau+d)^2 \wp(z,\tau).
\end{equation}
Differentiating Eq.(\ref{3eq9}) twice leads to the KdV nonlinear
differential equation of soliton physics in a time-independent way
\begin{equation}
\label{3eq12} \wp(z)'''=12\wp(z) \wp(z)'.
\end{equation}

 Now we consider the radiation-dominated Friedmann equation
\begin{equation}
\label{3eq13} \dot R^2=\frac{C_2}{R^2} +\frac{\Lambda}{3}R^2 -1.
\end{equation}
In this case the solution is given by~\cite{AAC}
\begin{equation}
\label{3eq14} R(v,\tilde
\tau)=\sqrt{\frac{3C_2}{12\wp(v+\epsilon,\tilde
\tau)+1}},~~t(v,\tilde \tau)=\frac{1}{2} \int R(v,\tilde \tau) dv
\end{equation}
where  $v(C_2,\Lambda)$ is the complex coordinate and $\tilde
\tau(C_2,\Lambda)$ is a modular parameter. These two describing a
new torus are  functions of both $C_2$ and $\Lambda$.  Here the
cubic invariants are given by
\begin{equation}
\label{3eq15} g_2=\frac{1}{12}-\frac{\Lambda
C_2}{12},~~g_3=\frac{1}{216}-\frac{\Lambda C_2}{144}.
\end{equation}
From Eqs.(\ref{3eq10}) and (\ref{3eq15}), if $\Lambda=0$, one
finds that discriminant is zero $(\Delta =0)$. The solutions to
this case are no longer given by the elliptic functions and do not
have modular properties. These were previously discussed in
Sec.3.1. Assuming a CFT  with $(L_0,c)$ on a torus, a partition
function with modular parameter $\tau$ can be introduced as
\begin{equation}
\label{3eq16} Z(\tau(C_1,\Lambda))={\rm Tr} q^{L_0-c/24},~~q=e^{2
\pi i \tau},
\end{equation}
where we suppress the $\bar \tau$-part for simplicity. Making use
of the modular properties of this partition function, we may find
the density of states and a two-dimensional Cardy formula for a
matter-dominated universe
\begin{equation}
\label{3eq17} S_{matter}= 2 \pi
\sqrt{\frac{c}{6}\Big(L_0-\frac{c}{24} \Big)}.
\end{equation}
Similarly, by assuming a CFT with $(\tilde L_0,\tilde c)$, we
expect to have  $S_{radiation}= 2 \pi \sqrt{\frac{\tilde
c}{6}\Big(\tilde L_0-\frac{\tilde c}{24} \Big)}$ for a
radiation-dominated universe, which is the same form as in
Eq.(\ref{3eq17}). These may lead to the chain connections:
Friedmann equation $\to$ Weierstrass equation ($\wp$) $\to$ torus
with $\tau$ $\to$ CFT partition function($Z(\tau(C_1,\Lambda))$)
$\to$ Cardy formula. However, we don't know exactly what kind of a
CFT is suitable for our purpose. Further, the cosmological
parameters of $\Lambda,C_1,C_2$ are used only for determining the
geometry of a torus itself. This presumed  mapping from the
Friedmann equation on $R^1 \times S^3$ into the Cardy formula on
torus ($T^2$) is not clearly justified. Actually we do not obtain
a direct definition of the quantities of $L_0(\tilde L_0),c(\tilde
c)$ appearing in the Cardy formula as a function of
$\Lambda,C_1,C_2$.
 
Consequently the existence of a Cardy formula from the solution to
the Friedmann equation is not clear and even if it is found, the
role of the cosmological constant $\Lambda$ always remains as a
modular parameter of torus. Also we note that the Verlinde's map
from the Friedmann equation to the Cardy-Verlinde formula  is
based on a CFT with a large central charge on 3-sphere of radius
$R$ ($S^3$) not torus ($T^2$).

\sect{Cosmological holographic bounds}
 
In this section we study two aspects of the cosmological term in
the holographic description of the universe : a look of entropy
($\Lambda \to S_{\rm \Lambda}$) and a look of energy ($\Lambda \to
E_{\rm \Lambda}$). In order to study the first aspect, we
introduce four holographic entropies
 which are necessary for making  the holographic description of a
  radiation-dominated universe with
  a positive cosmological constant~\cite{Verl,CM1}
 \footnote{Although Bousso argued that
a cosmological constant did not carry a genuine matter
entropy~\cite{Bous2},
 there is no contradiction to introducing
the geometric entropy. $S_{\rm \Lambda}$ was constructed by
analogy of the Hubble entropy $S_{\rm H }$. But $S_{\rm \Lambda}$
is closely related to the maximal de Sitter entropy of $S_{\rm
dS}$. Explicitly this is given by the Bekenstein-Hawking entropy
of the de Sitter cosmological horizon ($(n-1) V_{\rm dS}/4G_{n+1}
l_{n+1} \approx S_{\rm dS}=A/4G_{n+1}$) times the number ($N_{\rm
dS}= V/V_{\rm dS}$) of de Sitter regions in the early universe. }:
\begin{eqnarray}
\label{4eq1}
 {\rm Bekenstein-Verlinde\ entropy}:&& S_{\rm
 BV}=\frac{2\pi}{n}ER,
   \nonumber \\
 {\rm Bekenstein-Hawking\ entropy}:&& S_{\rm
 BH}=(n-1)\frac{V}{4G_{n+1}R},
    \nonumber \\
  {\rm Hubble\ entropy}:&& S_{\rm H}=(n-1)\frac{HV}{4G_{n+1}},
       \nonumber \\
  {\rm Cosmological \ entropy}:&& S_{\rm \Lambda}=
  (n-1)\frac{V}{4G_{n+1}l_{n+1}}.
\end{eqnarray}
$S_{\rm BV} \le S_{\rm BH}$ is supposed to hold for a weakly
self-gravitating universe ($HR \le 1$), while
 $S_{\rm BV} \ge S_{\rm BH}$ works when the universe is in the
  strongly self-gravitating
phase ($HR \ge 1$).
 It is interesting to
note that for  $HR=Hl_{n+1}=1$, one finds that four entropies are
identical:
 $S_{\rm BV}= S_{\rm BH}=S_{\rm H}=S_{\rm \Lambda}$.
In the holographic approach, it is useful to consider $S_{\rm BV}$
not really as an entropy but rather as the energy. And the
remaining three belong to the geometric entropy.   Then the first
Friedmann equation (\ref{1eq2}) can be expressed in terms of the
above four entropies as
\begin{equation}
 \label{4eq2} S^2_{\rm H} +(S_{\rm BV}-S_{\rm BH})^2
 =S^2_{\rm BV} +S^2_{\rm \Lambda}.
\end{equation}

In this section we no longer consider the matter-dominated case
because  one cannot transform the first Friedmann equation
 into the cosmological Cardy-Verlinde formula to
find the cosmological holographic bounds. This is mainly because
its energy-density  is given by $\rho_m=\rho_{m0}/R^3$  and the
solution $R(t)$ is expressed as a cycloid. In this case, the above
entropies are not suitable for representing the cosmological
holographic  bounds. On the other hand, for a radiation-dominated
case, there does not exist any difficulty in representing the
Friedmann equation in terms of the above four entropies. In this
case we have $\rho_r=\rho_{r0}/R^4$ and $R(t)$ is expressed as the
semicircle. As is shown in the footnote 3, the same nature of
entropy solution can be obtained by substitution : $R(\eta)
\leftrightarrow S_{\rm H}(\eta),~t(\eta) \leftrightarrow S_{\rm
BH}(\eta),~\sqrt{C_2} \leftrightarrow S_{\rm BV}$.  This is why we
will study a radiation-dominated universe without/with the
cosmological constant.
 
\subsection{Radiation-dominated universe without a cosmological
constant}

We start with $\Lambda_{n+1}=0$ case because this case gives us a
concrete relation. We  define a quantity $E_{\rm BH}$ which
corresponds to energy needed to form a universe-size black hole :
$ S_{\rm BH}=(n-1)V/4G_{n+1}R \equiv 2\pi E_{\rm BH} R/n $. With
this quantity, the Friedmann equations (\ref{1eq2}) and
(\ref{1eq3}) can be further cast to the cosmological
entropy-energy relation (cosmological Cardy-Verlinde formula) and
the cosmological Smarr formula respectively
\begin{eqnarray}
\label{4eq3}
 && S_{\rm H}=\frac{2\pi R}{n}\sqrt{E_{\rm
BH}(2E-E_{\rm
BH})}, \nonumber \\
&& E_{\rm BH}=n(E+pV -T_{\rm H} S_{\rm H}),
\end{eqnarray}
where the Hubble temperature ($T_{\rm H}$) as the minimum
temperature during the strongly gravitating phase is given by
$T_{\rm H}=-\frac{\dot H}{ 2\pi H}$. These are another
representation of the two Friedmann equations expressed in terms
of holographic quantities. On the other hand, we propose that the
entropy of a radiation-matter and its Casimir energy can be
described by the Cardy-Verlinde formula and the Smarr formula
respectively
\begin{eqnarray}
\label{4eq4} && S =\frac{2\pi R}{n}\sqrt{E_c(2E-E_c)},
  \nonumber \\
&& E_c=n(E+pV -T S).
\end{eqnarray}
 The first denotes  the entropy-energy relation, where $S$ is
 the entropy of a CFT-like radiation living on
an $n$-dimensional sphere with radius $R$ ($S^n$) and  $E$ is the
total energy of the CFT. Further the second represents the
relation between  a non-extensive part of the total energy
(Casimir energy) and thermodynamic quantities.  Here $E_c$ and $T$
stand for the Casimir energy of the system and the temperature of
radiation with $\omega=1/3$. Actually the above equations
correspond to thermodynamic relations for the CFT-radiation which
are originally independent of the geometric Friedmann equations.
Suppose that the entropy of radiation in the FRW universe can be
described by the Cardy-Verlinde formula. Then comparing
(\ref{4eq3}) with (\ref{4eq4}), one  finds that if $E_{\rm
BH}=E_c$, then $S_{\rm H}=S$ and $T_{\rm H}=T$. At this stage we
introduce the Hubble bound for entropy, temperature and Casimir
energy~\cite{Verl}
 \begin{equation}
 \label{4eq5}
 S \le S_{\rm H},~~ T \ge T_{\rm H},~~~E_c \le E_{\rm BH},
  ~~{\rm for}~ HR \ge 1
 \end{equation}
which shows  inequalities between geometric quantities and matter
contents. The Hubble entropy bound can be saturated by the entropy
of a radiation-matter filling  the universe when its Casimir
energy $E_c$ is
 enough to form a universe-size black hole.
If this happens, equations (\ref{4eq3}) and (\ref{4eq4}) coincide
exactly.
  This implies that the first Friedmann equation
somehow knows the entropy formula of a square-root form for a
radiation-matter filling the universe. As an  example, one
considers a moving brane universe in the background of the
five-dimensional Schwarzschild-AdS black hole. Savonije and
Verlinde~\cite{SV} found that when this  brane crosses the black
hole horizon, the Hubble entropy bound  is saturated by the
entropy of black hole(=the entropy of the  CFT-radiation). At this
moment the Hubble temperature and energy ($T_{\rm H},E_{\rm BH}$)
equal to the temperature and Casimir energy ($T,E_c$) of the
CFT-radiation dual to the AdS black hole respectively.
 
\subsection{Radiation-dominated universe  with a
positive cosmological constant: a look of entropy}
 
For a radiation-dominated universe with $\Lambda_{n+1} \not=0$, we
have to introduce the
 cosmological D-entropy $S_{\rm D}$  and D-temperature $T_{\rm D}$
 as~\cite{CM1}
\begin{equation}
\label{4eq6} S_{\rm D} =\sqrt{|S^2_{\rm H}-S^2_{\rm \Lambda}|}
,~~T_{\rm D}=- \frac{ \dot H} { 2 \pi \sqrt{|H^2-1/l^2_{n+1}|}}.
\end{equation}
We note that the cosmological D-entropy $S_{\rm D}$ is constructed
by analogy of the static D-bound\footnote{Suppose M is
asymptotically de Sitter space. Then the entropy of matter in M is
bounded by  the difference (D) between the entropy of exact de
Sitter space and the Bekenstein-Hawking entropy of the apparent
cosmological horizon in M of asymptotically de Sitter space.}.
$T_D$ is the lower bound of the  temperature during the strongly
gravitating phase with a positive cosmological constant.  Here we
insist that the first three entropies appeared in Eq.(\ref{4eq1})
are still applicable  for describing the radiation-dominated
universe with $\Lambda_{n+1} \not=0$ without any modification. As
a check point, one can recover the radiation-dominated universe
without a cosmological constant, as $\Lambda_{n+1} \to 0$ :
\begin{equation}
\label{4eq7} S_{\rm \Lambda} \to 0, ~~ S_{\rm D} \to S_{\rm H}
,~~T_{\rm D} \to T_{\rm H}.
\end{equation}
Using $S_{\rm D}$, one finds from Eq.(\ref{4eq2}) the entropy
relation
\begin{equation}
 \label{4eq8} S^2_{\rm D} +(S_{\rm BV}-S_{\rm BH})^2 =S^2_{\rm BV}.
\end{equation}
which is the same relation for $\Lambda_{n+1}=0$ case in the
footnote 3. Hence  $S_{\rm BV}$ is constant, $S_{\rm D}=S_{\rm BV}
\sin \eta$ and $S_{\rm BH}=S_{\rm BV}(1-\cos \eta)$ change with
time. Then Eqs.(\ref{4eq8}) and (\ref{1eq3}) can be rewritten as
the cosmological Cardy-Verlinde and cosmological Smarr formulas
\begin{eqnarray}
\label{4eq9}
 && S_{\rm D}=\frac{2\pi R}{n}\sqrt{E_{\rm
BH}(2E-E_{\rm
BH})}, \nonumber \\
&& E_{\rm BH}=n(E+pV -T_{\rm D} S_{\rm D}),
\end{eqnarray}
 while the entropy and  Casimir energy  of
the CFT-radiation can be expressed as
\begin{eqnarray}
\label{4eq10} && S =\frac{2\pi R}{n}\sqrt{E_c(2E-E_c)},
  \nonumber \\
&& E_c=n(E+pV -T S).
\end{eqnarray}
As is shown in Eq.(\ref{4eq7}), the cosmological D-entropy plays
the same role as the Hubble entropy does in the case without a
positive cosmological constant. That is, it is also a geometric
entropy during the strongly gravitating phase with a positive
cosmological constant.
 
Now we are in a position to see how the entropy bounds are changed
here. The first Friedmann equation in Eq.(\ref{1eq2})  can be
rewritten as
\begin{equation}
\label{4eq11} (HR)^2-\frac{R^2}{l_{n+1}^2}= 2 \frac{S_{\rm
BV}}{S_{\rm BH}} -1.
\end{equation}
Using this relation, in  case of $\Lambda_{n+1}=0$, one finds that
$HR \ge 1 \to S_{\rm BV} \ge S_{\rm BH}$, while $HR \le 1 \to
S_{\rm BV} \le S_{\rm BH}$. Hence this leads to the Hubble entropy
bound of $S \le S_{\rm H}$ for $HR \ge 1$, whereas the
Bekenstein-Verlinde entropy bound of  $S \le S_{\rm BV}$ for $HR
\le 1$. For   $ \Lambda_{n+1} \not=0$, it is shown that $(HR)^2
-\frac{R^2}{l_{n+1}^2} \ge 1
  \to S_{\rm BV} \ge
S_{\rm BH}$, while $ (HR)^2 - \frac{R^2}{l_{n+1}^2}\le 1  \to
S_{\rm BV} \le S_{\rm BH}$. Thus  this leads to the cosmological
D-bound for entropy, temperature, and Casimir energy for the
strongly gravitating phase:
\begin{equation}
\label{4eq12} S \le S_{\rm D},~~T \ge T_{\rm D},~~E_c \le E_{\rm
BH},~~ {\rm for},~~ HR \ge \sqrt {1+ \frac{R^2}{l_{n+1}^2}},
\end{equation}
whereas the Bekenstein-Verlinde entropy bound is found for the
weakly gravitating phase:
\begin{equation}
\label{4eq13} S \le S_{\rm BV},~~{\rm for}~~ HR \le \sqrt {1+
\frac{R^2}{l_{n+1}^2}}.
\end{equation}
When the  cosmological D-entropy bound is saturated by the entropy
$S$ of a CFT-radiation,  equations (\ref{4eq9})  and (\ref{4eq10})
 coincide, just like the case without the
cosmological constant. We note that one cannot find the relation
of $S_{\rm D}= S_{\rm BV}=S_{\rm BH}$ for $HR=1$, unless
$\Lambda_{n+1}=0$.
 
\subsection{Radiation-dominated universe  with a
positive cosmological constant : a look of energy}
 
If the cosmological term in Eq.(\ref{1eq2}) takes a closer look of
the potential energy term, then we can incorporate this into the
Bekenstein-Verlinde entropy. Noting that the Bekenstein-Verlinde
entropy $S_{\rm BV}$ is really considered as an energy, the
cosmological term appears in an additive form of energy in the
cosmological Cardy-Verlinde formula Eq.(\ref{4eq3}) without
introducing $S_{\rm D}$. Introducing the corresponding energy
$E_{\rm \Lambda}=\frac{\Lambda_{n+1} V}{8 \pi G_{n+1}}$
(equivalently, the last term in Eq.(\ref{1eq2}) is given by
$\frac{1}{l^2_{n+1}}=\frac{16 \pi G_{n+1}}{n(n-1)}\frac{E_{\rm
\Lambda}}{V}$), the Friedmann equations take the form instead of
Eq.(\ref{4eq8})~\cite{Youm}
\begin{eqnarray}
\label{4eq14}
 && S_{\rm H}=\frac{2\pi R}{n}\sqrt{E_{\rm
BH}\Big[2(E+E_{\rm \Lambda})-E_{\rm
BH}\Big]}, \nonumber \\
&& E_{\rm BH}=n(E+pV -T_{\rm H} S_{\rm H}).
\end{eqnarray}
On the other hand, the entropy and  Casimir energy  of the
CFT-radiation remains unchanged as
\begin{eqnarray}
\label{4eq15} && S =\frac{2\pi R}{n}\sqrt{E_c(2E-E_c)},
  \nonumber \\
&& E_c=n(E+pV -T S).
\end{eqnarray}
The above two forms do not resemble each other, because on the CFT
side, it is hard to incorporate the bulk cosmological term into
the Cardy-Verlinde formula. In terms of naive power counting, the
vacuum energy (cosmological term) corresponds to a relevant
operator in CFT. And this leads to power divergences. In this case
it is not easy to obtain the cosmological holographic bounds like
Eqs. (\ref{4eq5}) and (\ref{4eq12}). Furthermore, introducing a
related entropy $\tilde S_{\rm \Lambda}= \frac{2 \pi R}{n} E_{\rm
\Lambda}$ like $S_{\rm BH}=\frac{2 \pi R}{n} E_{\rm BH}$, the
entropy relation in Eq.(\ref{4eq2}) is changed into an ugly form
as
\begin{equation}
 \label{4eq16} S^2_{\rm H} +(S_{\rm BV}+\tilde S_{\rm
\Lambda}-S_{\rm BH})^2 =(S_{\rm BV} + \tilde S_{\rm \Lambda})^2.
\end{equation}
Here  $S_{\rm BV} + \tilde S_{\rm \Lambda}$ does not remain
constant during the cosmological evolution unlike
$\Lambda_{n+1}=0$ case shown in the footnote 3 and Eq.(\ref{4eq8})
for $\Lambda_{n+1} \not=0$ case.
 
\sect{Discussion} In this work we discuss the role of the
cosmological constant in the early universe. Especially for the a
radiation-dominated universe $\rho_r=\rho_{r0}/R^4$  with a
positive cosmological constant $\Lambda$, we confirm the
cosmological holographic bounds Eq.(\ref{4eq12}) if the
cosmological constant is considered as an entropy ($\Lambda \to
S_{\rm \Lambda}$). Here the entropy concept originates from the
Hubble entropy $S_{\rm H}$ which plays a crucial role in
establishing the cosmological holographic principle in the
radiation-dominated universe. We note here that the two entropies
$S_{\rm H}$ and $S_{\rm \Lambda}$ are regarded as the geometric
entropy but not the genuine matter entropy like $S$ for a
CFT-radiation matter.
 
Taking a genuine view of energy ($\Lambda \to E_{\rm \Lambda}$),
one cannot establish the cosmological holographic bounds in the
early universe. For the matter-dominated universe without/with a
positive cosmological constant, we cannot achieve the cosmological
holographic bounds because of its energy density nature with
$\rho_m=\rho_{m0}/R^3$. Further, for the pure de Sitter case
without any radiation, one cannot derive the cosmological
holographic bounds~\cite{myung1}.
 
Finally, considering the cosmological constant term as a candidate
of dark energy in the present universe, its role of the geometric
entropy in the holographic description of the early universe
emerges as an opposite one. If this view is correct, our work
implies  a duality of the cosmological constant :  (geometric)
entropy in the early universe and (dark) energy in the present
universe.
\section*{Acknowledgment}
We thank  C.-R.Cai for helpful discussions. This work was
supported in part by KOSEF, Project Number: R02-2002-000-00028-0.

\end{document}